\documentclass[reprint,aip,apl]{revtex4-1}
\usepackage{graphicx}
\usepackage{upgreek}
\usepackage{dcolumn}
\usepackage{bm}% bold math

\usepackage{colordvi}
\begin{document}
\title{Long-lived spin polarization in n-doped MoSe$_2$ monolayers}
\author{M.\ Schwemmer}
\affiliation{Institut f\"ur Experimentelle und Angewandte Physik, Universit\"at
Regensburg, D-93040 Regensburg, Germany}
\author{P.\ Nagler}
\affiliation{Institut f\"ur Experimentelle und Angewandte Physik, Universit\"at
Regensburg, D-93040 Regensburg, Germany}
\author{A.\ Hanninger}
\affiliation{Institut f\"ur Experimentelle und Angewandte Physik,
Universit\"at Regensburg, D-93040 Regensburg, Germany}
\author{C.\ Sch\"uller}
\affiliation{Institut f\"ur Experimentelle und Angewandte Physik,
Universit\"at Regensburg, D-93040 Regensburg, Germany}
\author{T.\ Korn}
\email{tobias.korn@physik.uni-regensburg.de}
\affiliation{Institut f\"ur Experimentelle und Angewandte Physik,
Universit\"at Regensburg, D-93040 Regensburg, Germany}
\date{\today}
\begin{abstract}
Transition metal dichalcogenide monolayers are highly interesting for potential valleytronic applications due to the coupling of spin and valley degrees of freedom and valley-selective excitonic transitions. However,  ultrafast recombination of excitons in these materials poses a natural limit for applications, so that a transfer of polarization to resident carriers is highly advantageous.  Here, we study the low-temperature spin-valley dynamics in nominally undoped and n-doped MoSe$_2$ monolayers using time-resolved Kerr rotation. In the n-doped MoSe$_2$, we find a long-lived component of the Kerr signal which we attribute to the spin polarization of  resident carriers. This component is absent in the nominally undoped MoSe$_2$. The long-lived spin polarization is stable under applied in-plane magnetic fields. Spatially resolved measurements allow us to determine an upper boundary for the electron spin diffusion constant in MoSe$_2$.
\end{abstract}
\maketitle
Transition metal dichalcogenide (TMDC) monolayers have a peculiar band structure~\cite{Mak2010,Splendiani2010}, in which spin and valley degrees of freedom are coupled, and the optical selection rules allow for valley-selective generation of excitons~\cite{Xiao2012}. An excitonic valley polarization can be read out optically via the circular polarization degree of the emitted photoluminescence (PL), and initial studies using continuous-wave excitation revealed a large steady-state polarization~\cite{Zeng2012,Mak2012}, which was shown to be stable against depolarization in large in-plane magnetic fields~\cite{Sallen2012}. While many early studies of valley physics focused on the naturally abundant MoS$_2$, synthetic TMDC crystals quickly garnered scientific attention due to their spectrally narrow PL emission~\cite{Tonndorf2013}, and the large tuning range of conduction- and valence-band spin splitting~\cite{Dery_FlexPhonons,Kormanyos2015,Dery_Excitons} afforded by the different combinations (or alloys~\cite{Mann2014,Urbaszek_MoWSe}) of the constituent elements Mo, W, S, Se, and Te. Time-resolved studies  of exciton recombination and valley dynamics showed that ultrafast radiative recombination~\cite{Korn2011,Lagarde2014,Poellmann2015,Marie16,Bayer16} is partially responsible for the large valley polarization  observed in experiments, as the exciton lifetime limits the time window for valley dephasing. However, the ultrafast excitonic recombination also severely limits potential applications of the valley degree of freedom in novel devices, so that a transfer of valley polarization to resident carriers, which was observed in several TMDCs~\cite{Yang2015a,Bushong_Arxiv}, is highly advantageous.  Among the TMDC family, MoSe$_2$ is characterized by a comparatively low optically induced valley polarization degree, which is only observable at all under near-resonant excitation conditions~\cite{Wang2015e,Hanbicki_SciRep, Plochocka_2Dmat17} and can be increased by modifying the recombination dynamics using coupling to photonic cavities~\cite{Lundt17}. To investigate the anomalously low valley polarization in MoSe$_2$, time-resolved studies of valley dynamics are needed.

Here, we directly compare the low-temperature spin-valley dynamics in nominally undoped and n-doped MoSe$_2$ monolayers using time-resolved Kerr rotation (TRKR). We find a long-lived component of the Kerr signal in the n-doped MoSe$_2$ which is absent in the nominally undoped sample. We attribute this to an optically induced spin polarization of  resident carriers. TRKR measurements in an applied in-plane magnetic field show no influence of the field on this spin polarization, indicating that it is stabilized by the large, valley-contrasting conduction-band spin splitting of MoSe$_2$.

The investigated samples are prepared from  bulk crystals (deliberately n-doped MoSe$_2$ supplied by 2D semiconductors and MoSe$_2$ without intentional doping and unspecified majority carrier type supplied by HQ graphene)  by mechanical exfoliation. For this, we use Gel-Film$^{TM}$ pieces (Gelpak) as an intermediate substrate, and suitable monolayer flakes are  then stamped onto the final substrate (silicon covered with a 285\,nm SiO$_2$ layer and pre-defined metal markers)  using an all-dry transfer technique~\cite{Castellanos2014}.    The investigated samples are identified as HQ (monolayer flake prepared from HQ graphene MoSe$_2$ bulk crystal) and n-2D (monolayer flake prepared from n-doped 2D semiconductors MoSe$_2$ bulk crystal) throughout the manuscript.
PL and time-resolved PL (TRPL) measurements are performed in a self-built microscope setup, details are described elsewhere~\cite{Nagler17_2Dmat}.  The time resolution of the streak camera used for the TRPL measurements is below 10~ps.
Details of the setup for the TRKR measurements are  described in Ref.~\onlinecite{Schwemmer16}. Briefly,  a mode-locked Ti:Sapphire laser emitting 100~fs pulses is used as the source for pump and probe beams. Band pass filters cut non-overlapping high-energy (pump) and low-energy (probe) regions out of the broad laser spectrum. A mechanical delay stage is used to define the time delay $\Delta t$ between pump and probe pulses. Pump and probe beams are collinearly focused via a 100x microscope objective onto the sample, yielding spot diameters of about 1.5~$\upmu$m  and excitation densities of about 9~$\textrm{kW}\textrm{cm}^{-2}$ (pump) and 27~$\textrm{kW}\textrm{cm}^{-2}$ (probe), respectively.    The pump pulses are circularly polarized to generate a spin-valley polarization, while the probe pulses are linearly polarized to probe this polarization via the magneto-optical Kerr effect. The pump beam is filtered out from the back-reflected light by a long-pass filter, and the induced ellipticity of  the reflected probe beam is detected by an optical bridge detector. Since we utilize a spectrally broad probe pulse with an FWHM of about 20~meV, measuring ellipticity avoids partial cancelation of the Kerr rotation signal for different spectral components of the probe, as the Kerr rotation angle changes its sign when the probe energy is tuned across the resonance frequency of an excitonic transition, while the ellipticity has a maximum at the resonance frequency~\cite{Glazov2012}.
Changing the pump-beam path via a piezo-electric mirror holder allows to move the pump spot laterally on the sample surface with respect to the probe spot.  All PL and TRKR measurements are performed at a nominal sample temperature of 4.5~K, with the sample mounted on the cold finger of a liquid-He-cooled cryostat. An electromagnet is integrated into the TRKR setup, so that magnetic fields can be applied in the sample plane.
\begin{figure}
\includegraphics[width=\linewidth]{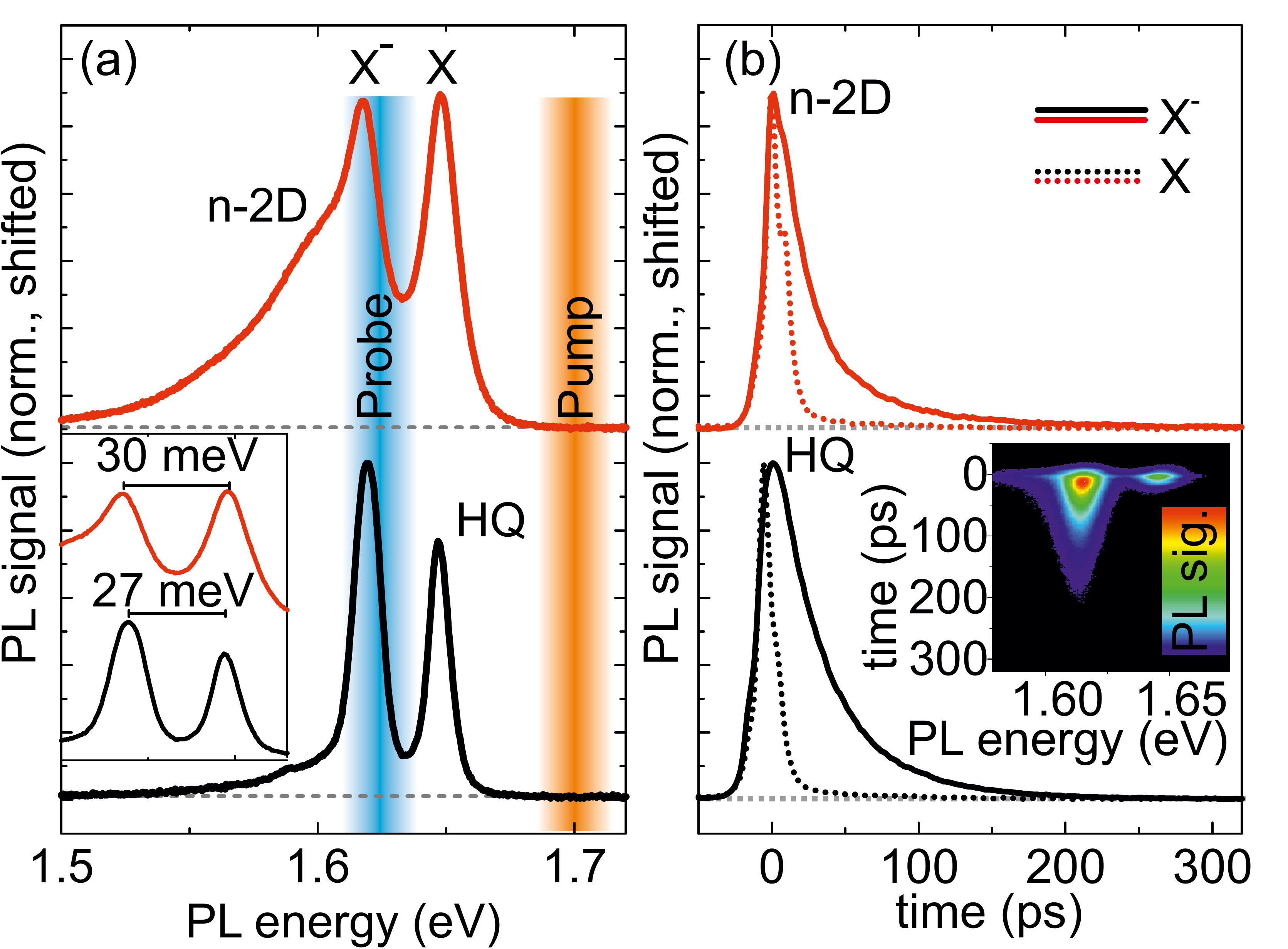}
\caption{\label{Fig1_PL_TRPL}
(a) Time-integrated PL spectra measured on n-2D (red lines) and HQ (black lines) samples at 4.5~K.  The shaded areas indicate the energetic positions and linewidth of the pump (orange) and probe (blue) pulses used in the TRKR measurements. The inset shows a small region of the same spectra around the exciton (X) and trion (X$^-$) peaks, with the energy differences between the X and (X$^-$) peaks indicated by the black scale bars. (b) Time-resolved PL traces measured on n-2D (red) and HQ  (black) samples at 4.5~K. Solid (dashed) lines indicate traces centered around trion (exciton) PL peaks. The inset shows a false-color plot of the time- and spectrally resolved PL of the HQ  sample measured at 4.5~K.}
\end{figure}

To characterize our samples, we perform time-integrated and time-resolved micro-PL measurements. Figure~\ref{Fig1_PL_TRPL}(a) shows the time-integrated PL spectra. In both samples, we clearly see well-resolved neutral exciton (X) and trion (X$^{-}$) peaks with comparable peak heights. However, we note that the trion peak in the n-2D sample has a pronounced, low-energy shoulder, which might stem from trions bound to ionized donor sites. We also observe a significantly larger exciton-trion energy splitting $\Delta XT$ in the n-2D sample. In TMDCs, $\Delta XT$ strongly depends on the Fermi energy, increasing with increasing carrier concentration~\cite{Mak2013,Chernikov_Tuning_WS2_PRL15,Plechinger2015a}, and therefore allows us to estimate differences in background carrier density $\Delta n_e$ of the two samples. Following the reasoning in Ref.~\onlinecite{Chernikov_Tuning_WS2_PRL15} and using a value of $m^{*}_{CB}=0.5 m_0$~\cite{Kormanyos2015} for the conduction-band electron mass in MoSe$_2$, we find $\Delta n_e = 7\cdot10^{11}$~cm$^{-2}$. We note that in order to determine the absolute background carrier density from PL spectra, the value of the exciton-trion energy splitting in the limit of low carrier concentration would be required. Nevertheless, the value of $\Delta n_e$ extracted from the PL spectra  serves as a lower boundary for the background carrier density.
Time-resolved PL measurements allow us to observe the  exciton and trion recombination dynamics. As the inset in Fig.~\ref{Fig1_PL_TRPL}(b) demonstrates, the characteristic times are very different: while the exciton PL emission decays on a timescale that is close to the system response time of our streak camera  the trion PL shows a markedly slower decay. This different behavior of exciton and trion is in qualitative agreement with previous TRPL studies on monolayer MoSe$_2$~\cite{Wang2015e,Marie16,Bayer16}.  To quantify the PL dynamics, we integrate the time-resolved PL intensity over a 10~meV wide window around the exciton and trion PL emission peaks, respectively. These TRPL traces are shown for both samples in Fig.~\ref{Fig1_PL_TRPL}(b). We find that the exciton PL emission decay can be described by a single exponential, with decay constants of 9~ps (HQ) and 10~ps (n-2D). By contrast, the trion PL emission is well-described by a biexponential fit, with fast components $\tau_1$=30~ps (HQ) and $\tau_1$=20~ps (n-2D), and a slow component $\tau_2$=75~ps for both samples. We note that the decay constants observed in our TRPL measurements are larger than those observed in previous studies, most likely due to the larger excess energy ( $>0.5$~eV) of the excitation laser used in our experiment.
\begin{figure}
\includegraphics[width=\linewidth]{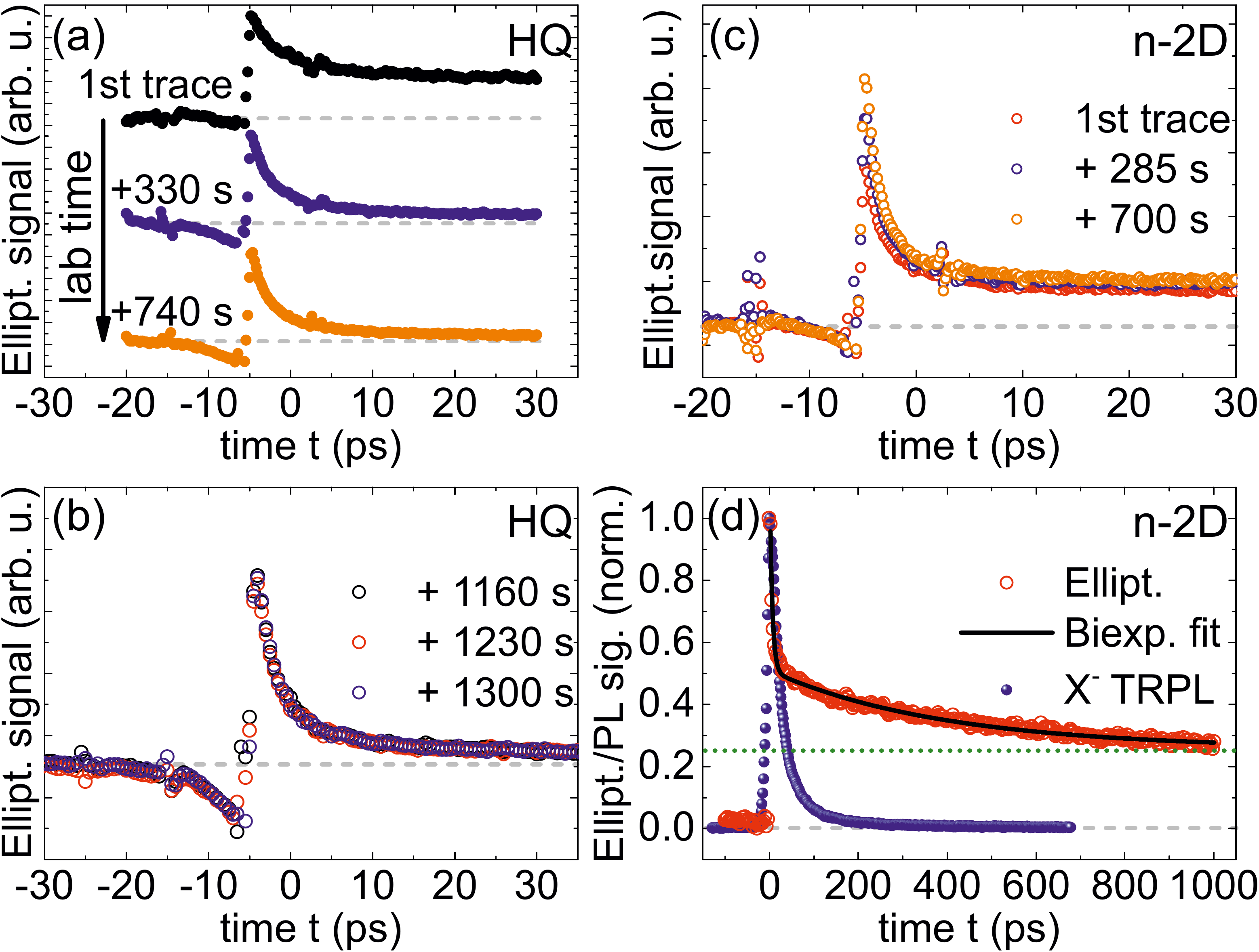}
\caption{\label{Fig2_Kerr_VGL}
(a) TRKR traces measured  HQ sample at 4.5~K as a function of lab time. The traces are shifted vertically with respect to each other. (b)Additional TRKR traces measured on same position of HQ sample as in (a), about 20 minutes after initial illumination of this sample position. (c) TRKR traces measured on  n-2D sample at 4.5~K as a function of lab time.  (d) Normalized TRKR (red) and trion TRPL (blue) traces measured on  n-2D sample at 4.5~K. The solid black line indicates a biexponential fit to the TRKR data. The dark green dotted line indicates the offset level of the biexponential fit.}
\end{figure}

To study the spin-valley dynamics in our samples, we turn to time-resolved Kerr rotation.   As indicated by the shaded areas in~\ref{Fig1_PL_TRPL}(a), our pump is centered around 1.7~eV, so that the excitation has an excess energy of about 50~meV with respect to the neutral exciton transition, while the probe is centered around 1.63~eV, close to the trion transition in both samples. We note that due to the large spectral width of the probe and the fixed energy windows available in the experiment (given by the band pass filters used to spectrally shape the probe), we cannot exclude that the measured ellipticity signal stems, in part, also from exciton valley polarization.

We first discuss the sample HQ. Figure~\ref{Fig2_Kerr_VGL}(a) shows a series of TRKR traces measured at 4.5~K. All measurements were taken on a fixed sample position, with the first trace measured immediately after the laser spot was moved to that position. For easy comparison, the TRKR traces have been shifted vertically. We clearly observe that the shape of the traces changes significantly as a function of lab time, corresponding to continuous illumination of the sample spot. While the initial trace shows a pronounced long-lived component that exceeds the measurement window shown in Fig.~\ref{Fig2_Kerr_VGL}(a), this component is suppressed in subsequent measurements. As we will discuss below, we associate this long-lived component with a spin polarization of resident carriers. Due to the high excitation density used in our experiment, the local carrier density within the HQ sample is reduced on a scale of minutes, most likely due to the desorption of adsorbates at the sample surface. Changes of the carrier density as a function of laser illumination have previously been observed in other TMDCs~\cite{Currie2015, Wurstbauer_photodoping}. Thus, the long-lived component due to resident carriers is suppressed. As indicated in Fig.~\ref{Fig2_Kerr_VGL}(b), the shape of the TRKR traces is stabilized after about 20~minutes of continuous illumination at a specific sample position. As the sample is moved to a new position, this behavior is reproduced (not shown), with the initial trace having a long-lived component that is suppressed in subsequent measurements. After stabilization, the TRKR traces are well-described by a biexponential fit function, with a fast component $t_1$=2~ps and a slow component $t_2$=11~ps, both of which are smaller than the decay constants observed in TRPL measurements for exciton and trion, indicating that exciton and trion recombination are not the limiting factors for the decay of the valley polarization tracked in the TRKR measurements. This rapid decay of the valley polarization is in good agreement with the low circular polarization degree of PL emission in MoSe$_2$ monolayers observed in previous studies~\cite{Wang2015e,Hanbicki_SciRep, Plochocka_2Dmat17}. Due to our excitation conditions providing substantial excess energy  and correspondingly large initial center-of-mass momentum to the exciton and trion states, the valley dephasing mechanism based on long-range exchange interaction~\cite{Glazov2014,Yu2014a} is highly efficient and compatible with the few-picosecond decay times we observe.
Remarkably, the TRKR traces measured on the n-2D sample show a very different behavior: as seen in Fig.~\ref{Fig2_Kerr_VGL}(c), the traces are very stable as a function of lab time, with the initial trace closely matching subsequent traces. All measurements show a substantial long-lived component, which significantly exceeds the measurement window of about 1~ns limited by the mechanical delay stage of our experiment, as shown in Fig.~\ref{Fig2_Kerr_VGL}(d). On the few-picosecond timescale, the decay of the signal is well-described by a biexponential decay function with a constant offset corresponding to the long-lived component. The fast and slow components of this fit yield values of $t_1$=1.3~ps and  $t_2$=6~ps, respectively, demonstrating faster dephasing than in the HQ sample, which might stem from increased scattering of excitons and trions with the large resident carrier density in this sample. On the long timescale shown in Fig.~\ref{Fig2_Kerr_VGL}(d), we also utilize a biexponential decay function with a constant offset to account for the very long-lived component that exceeds our measurement window. While the fast $t_1$ dynamics observed on the few-ps timescale are not visible in this dataset, which was recorded with a 3~ps step size, we find a similar decay constant $t_2$=6~ps, and an additional, slower component of $t_3$=395~ps. The constant offset amounts to about 25~percent of the initial TRKR amplitude. Using the noise level of the TRKR signal before arrival of a pump pulse as a lower boundary for the TRKR signal remaining after 12~ns (corresponding to the repetition rate of the laser), we can determine an upper limit for the lifetime of the Kerr signal exceeding the time window available in our experiment, and we find a value of $t_F$=3.6~ns.
Since the lowest-energy excitonic state in MoSe$_2$ is bright, in contrast to the tungsten-based materials, where dark excitons and trions can contribute to a long-lived valley polarization~\cite{Plechinger16,Plechinger17,Beschoten_Arxiv}, we associate the long-lived component with a spin polarization of the resident carriers in the deliberately n-doped n-2D sample. This assertion is supported by the absence of such a long-lived component in the HQ sample with its lower carrier density. For these resident carriers, the long-range exchange mechanism responsible for exciton valley dephasing is absent, and naturally, the PL lifetime does not pose an upper limit for the spin polarization of resident carriers, as the direct comparison of PL dynamics and Kerr rotation trace in Fig.~\ref{Fig2_Kerr_VGL}(d) confirms. The lifetime we observe is in agreement with calculations for the electron spin lifetime in MoSe$_2$ at low temperatures~\cite{Dery_FlexPhonons}.
In order to initialize a resident spin polarization using optical excitation, a transfer of polarization from excitons to resident carriers has to occur, in which a subset of the optically generated holes dephases during exciton formation, so that unpolarized excitons are formed using the resident carrier reservoir, while a larger subset of the optically oriented electrons retains its polarization. Such an asymmetric dephasing may be driven by different electron and hole spin-flip scattering rates with the resident carriers, or by different spin dephasing rates during energy relaxation in the conduction and valence bands, which are common, e.g, in GaAs-based heterostructures~\cite{Kugler_nonresonant}.

\begin{figure}
\includegraphics[width=\linewidth]{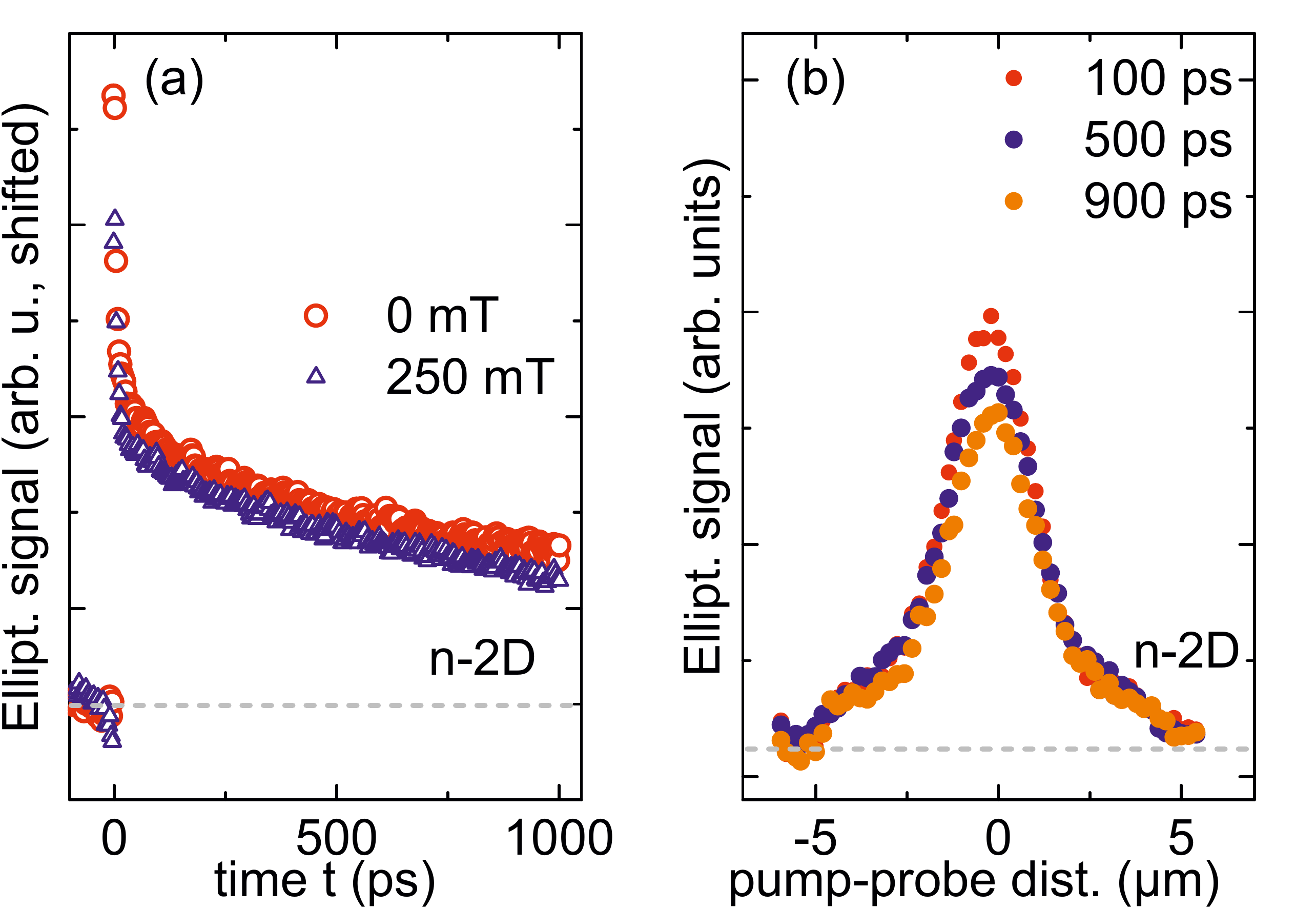}
\caption{\label{Fig3_Mag_LineScan}
(a) TRKR traces measured on n-2D sample at 4.5~K with (blue triangles) and without (red circles) an applied in-plane magnetic field of 250~mT. (b)TRKR line scans measured on n-2D sample at 4.5~K for different delays between pump and probe pulse.}
\end{figure}
To explore the stability of this resident spin polarization, we perform measurements in an applied in-plane magnetic field of 250~mT (the maximum field available in our electromagnet). As Fig.~\ref{Fig3_Mag_LineScan}(a) shows, there is neither any discernible spin precession observable, nor is there a significant change of the decay of the Kerr signal as compared to the zero-field measurement under similar experimental conditions. This result is in contrast to the resident carrier spin dynamics in the related material MoS$_2$, where a significant increase of the decay rate with magnetic field, combined with an oscillatory behavior, was observed~\cite{Yang2015a}. This qualitative difference stems from the larger conduction-band spin splitting of MoSe$_2$ (20~meV compared to 3~meV in MoS$_2$~\cite{Kormanyos2015}), which effectively acts as a large, valley-dependent out-of-plane magnetic field, suppressing spin precession about the applied in-plane magnetic field and stabilizing the spin polarization. A qualitatively similar behavior was also reported for resident spin polarization in WS$_2$~\cite{Bushong_Arxiv}.

Finally, we perform spatially resolved experiments, where the pump beam is scanned across the sample surface along a line, while the position of the probe beam is kept fixed. Figure~\ref{Fig3_Mag_LineScan}(b) shows the Kerr signal as a function of pump-probe beam distance for 3 different time delays between pump and probe. Naturally, the signal is maximum for overlapping pump and probe beams and decays with increasing distance, following a convolution of pump and probe beam spatial profiles. We find that the absolute signal amplitude decreases with increasing time delay, indicating the decay of the spin polarization. However, the width of the curve (extracted via fitting a Gaussian to the data) does not increase significantly as a function of time delay, indicating that  spin diffusion of resident carriers is below our detection limit. Using the spatial resolution of our setup and the available time window, we can estimate an upper boundary for the spin diffusion constant of 15~$\frac{\textrm{cm}^2}{\textrm{s}}$.

In conclusion, we have investigated the spin-valley dynamics in different MoSe$_2$ monolayer samples using time-resolved Kerr rotation. In an intentionally n-doped sample, we observe a long-lived component of the Kerr signal which is absent in a nominally undoped MoSe$_2$ sample. We attribute this  signal to a spin polarization of the resident electrons, which is established via optical excitation and subsequent transfer of polarization to resident carriers. Spatially resolved measurements allow us to determine an upper boundary for the electron spin diffusion constant. The long-lived electron spin polarization is stable against in-plane magnetic fields due to the large conduction-band spin splitting of MoSe$_2$, making n-doped MoSe$_2$ potentially interesting for valleytronic devices.

We gratefully acknowledge financial support by the Deutsche Forschungsgemeinschaft via GRK 1570, KO3612/1-1 and SFB689.
\end{document}